# Detailed Performance Loss Analysis of Silicon Solar Cells using High-Throughput Metrology Methods

Mohammad Jobayer Hossain[1], Geoffrey Gregory[2], Hardik Patel[2], Siyu Guo[3], Eric J. Schneller[2,3], Andrew M. Gabor[4], Zhihao Yang[5], Adrienne L. Blum[6], Kristopher O. Davis[1,2,3]

[1]CREOL, the College of Optics and Photonics, University of Central Florida, Orlando, FL, USA
[2]Department of Materials Science and Engineering, University of Central Florida, Orlando, FL, USA
[3]Florida Solar Energy Center, University of Central Florida, Cocoa, FL, USA
[4]BrightSpot Automation, Westford, MA, USA
[5]School of Materials Science and Energy Engineering, Foshan University, Guangdong, China
[6] Sinton Instruments, Boulder, CO, USA

*Abstract* — In this work, novel, high-throughput metrology methods are used to perform a detailed performance loss analysis of ≈400 industrial crystalline silicon solar cells, all coming from the same production line. The characterization sequence includes a non-destructive transfer length method (TLM) measurement technique featuring circular TLM structures hidden within the busbar region of the cells. It also includes a very fast external quantum efficiency and reflectance measurement technique. More traditional measurements, like illuminated current-voltage, Suns-$V_{OC}$, and photoluminescence imaging are also used to carry out the loss analysis. The variance of the individual loss parameters and their impact on cell performance are investigated and quantified for this large group of industrial solar cells. Some important correlations between the measured loss parameters are found. The nature of these distributions and correlations provide important insights about loss mechanisms in a cell and help prioritize efforts to optimize the performance of the production line.

## I. Introduction

Analyzing and quantifying the various energy conversion losses occurring in photovoltaic (PV) cells and modules is fundamental to better understanding how these devices behave and engineering them to be better (e.g., more efficient, less expensive, more durable). Performance loss analysis normally revolves around decoupling loss mechanisms (e.g., optical, recombination, resistive) and quantifying their influence on cell performance. Due to the time consuming nature of the measurements involved, a detailed performance loss analysis is typically only carried out on small groups of solar cells, normally just in R&D settings [1]. However, with the advent of new high-throughput metrology techniques, it has become more convenient to perform these types of studies on larger groups of cells. In this work, we've carried out five different measurement techniques on ≈400 industrial crystalline silicon (c-Si) solar cells, all from the same production line, and will present a detailed performance loss analysis on this statistically relevant group of cells. The five measurement techniques include: (1) illuminated *I-V* at standard test conditions, a common method used to test and bin cells following their fabrication [2]; (2) Suns-$V_{OC}$ [3]; (3) photoluminescence (PL) imaging [4]; (4) high-speed quantum efficiency and reflectance spectroscopy [5]; and (5) non-destructive transfer length method (TLM) measurements [6].

## II. Experimental Details and Analysis Techniques

The c-Si solar cells used in this work are industrial-sized, multicrystalline silicon cells featuring five busbars. These are industry standard Al-BSF (aluminum back surface field) cells, featuring isotropic texturing, a $SiN_x$ anti-reflection coating (ARC), phosphorus-doped $n^+$ emitter on the front, screen-printed Ag front contacts, and a full area screen-printed Al rear contact that forms the $p^+$ BSF upon firing. The only difference in these cells are the cTLM structures hidden within the busbars, which don't influence the cell processing or performance.

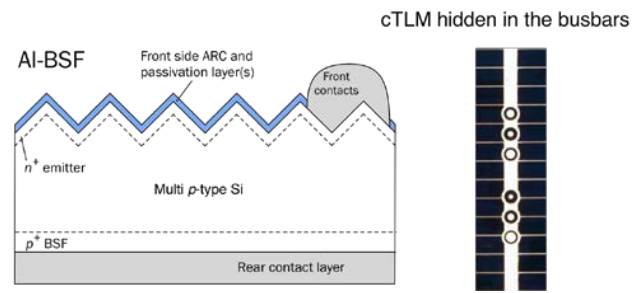

Fig. 1. (Left) Illustration of the cell architecture used in this study and (right) image of the cTLM structures hidden within the busbars of the cells.

Illuminated *I-V* curve measurements under standard test conditions and Suns-$V_{OC}$ measurements were performed using Sinton Instruments FCT-750 in-line cell tester. The FCT-750 is a production cell tester capable of measuring cells at line speeds of 3600 units per hour. This cell tester is unique in that it reports conventional cell test parameters ($I_{SC}$, $V_{OC}$, $R_S$, $R_{SH}$, power, efficiency, and fill factor) and supplements these parameters with an advanced Suns-$V_{OC}$ analysis and substrate doping measurement. The Suns-$V_{OC}$ analysis allows for a true $R_S$ measurement, pseudo $I$-$V$ parameter measurement without the effects of $R_S$, and carrier lifetime data. Open-circuit PL images were obtained with a BT Imaging *LIS-R1*. The imaging was carried out at two injection levels, 0.1 suns and 1sun, to obtain the spatial distribution of $V_{OC}$ over the cells. The laser in *LIS-R1* shines the sample with 808 nm light. PL emission occurs due to the radiative recombination in the sample. The broad emission spectrum is filtered by a long wavelength pass filter having a cutoff wavelength of 920 nm before it reaches the detector. It makes sure that the part of the laser excitation reflected from the sample does not reach the detector. 3.01•$10^{17}$ cm$^{-2}$s$^{-1}$ and 3.01•$10^{16}$ cm$^{-2}$s$^{-1}$ photon flux were assumed for 1 sun and 0.1 sun, respectively. The experiment was carried out at room temperature. A spatial open-circuit voltage image $V_{xy}$ can be obtained from the PL intensity image ($I_{Hxy}$) using the following equation [4]:

$$V_{xy} = V_T \cdot \log\left(\frac{I_{Hxy} - B_{xy} \cdot I_H}{C_{xy}}\right) \quad (1)$$

Here $V_T$ is the thermal voltage. $I_H$ is the illumination intensity at which the PL image $I_{Hxy}$ was taken. $B_{xy}$ is a called a background calibration constant which accounts for diffusion limited carriers. It is generally derived from an additional short-circuit current image. However, since $B_{xy}$ is negligible in comparison to $I_{Hxy}$, it has been ignored for simplicity. $C_{xy}$ is a calibration constant independent of electrical bias and illumination conditions. It can be expressed as,

$$C_{xy} = I_{Lxy} \cdot \exp\left(\frac{V_{OC}}{V_T}\right) \quad (2)$$

Here $I_{Lxy}$ is another open-circuit PL image at a lower illumination intensity (0.1 sun for this work). $V_{OC}$ is the measured open-circuit voltage.

The high-speed EQE and reflectance measurements were performed with a customized Tau Science *FlashQE* system featuring an *xy*-gantry and integrating sphere. The FlashQE syatem has 41 LEDs whose emission ranges from 365 nm to 1280 nm with nonuniform spacing among them. The integrating sphere diffuses the light from all the LEDs and combines them into a spot size of 4 mm diameter. The LEDs turn ON and OFF at different frequencies (KHz range). Their individual contribution to the total photogenerated current is decoupled by taking its Fourier tansform and correlating with the corresponding ON-OFF frequency. In this case, the EQE and reflectance were measured on a 10•10 grid for each cell (i.e., 100 points on each cell). Measurement at each of these points takes approximately 1 second. The measurement locations were set to avoid the busbars and had the same amount of physical shading fraction for all the measurements in a cell. Different loss components of the short-circuit current density ($J_{SC}$) [7] were extracted using the method outlined in [5], including: (a) front surface reflectance ($J_{R-f}$); (b) escape reflectance ($J_{R-esc}$) ; (c) absorption in the anti-reflection coating and loss in the emitter ($J_{loss-e}$); and (d) loss in the bulk and rear, or base ($J_{loss-b}$).

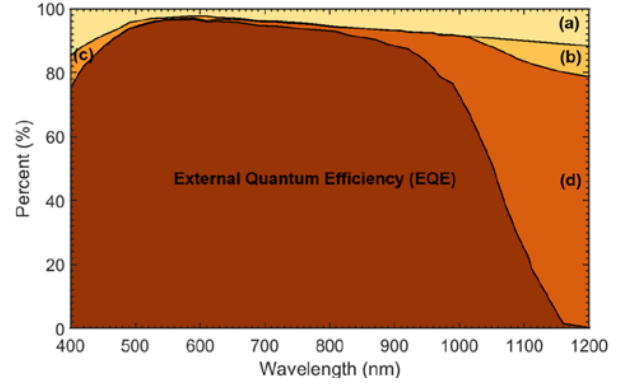

Fig. 2. EQE and reflectance spectrum highlighting various $J_{SC}$ losses (a) front surface reflectance, (b) escape reflectance (c) emitter loss (d) bulk and rear surface loss.

In the FlashQE measurement, light from the integrating sphere shines a spot on the solar cell. Part of it gets reflected from the front surface and gridlines. Busbars are avoided in the measurement process; therefor there is no reflection from them. The rest of the light enters into the cell and starts getting absorbed there. However, longer wavelength lights find it difficult to get absorbed. They are reflected from the rear side metallization and reach the detector escaping the cell. This kind of reflectance is called escape reflectance. Since escape reflectance occurs only in the longer wavelengths, the total reflection (sum of front and escape reflectance) starts increasing rapidly after certain point in longer wavelength region. Therefore a linear trendline beyond that point (≈1020 nm in this study) represents the front reflection in addition to all the reflections before that point. The difference between total and front reflection provides escape reflectance. The wavelength with minimum reflection tells about the thickness of the SiN$_x$ ARC as in equation 3. Considering first order reflection the thickness of ARC,

$$t = \frac{\lambda_{min}}{4n} \quad (3)$$

Here, $\lambda_{min}$ is the wavelength corresponding to minimum reflection, *n* is the refractive index of SiN$_x$. Since the absorption in the antireflection coating is negligible, the internal quantum efficiency (IQE) can be calculated from the measured value of EQE and reflectance,

$$IQE(\lambda) = \frac{EQE(\lambda)}{1 - R_{total}} \quad (4)$$

Apart from the losses due to reflection and shading, parasitic absorptions also occur in the bulk, emitter and the rear surface of the solar cell. Emitter losses occurs in the short wavelengths when bulk and rear losses occur mostly in the long wavelengths. The losses in the emitter can be modelled as a hypothetical dead layer thickness $W_{de}$. With this assumption, IQE can be expressed as,

$$IQE(\lambda) = \frac{1}{k}\exp\left(-\frac{W_{de}}{L_a(\lambda)}\right)\frac{1}{1+\frac{L_a(\lambda)}{L_{eff}}} \quad (5)$$

Here, $L_{eff}$ and $L_a$ are the effective diffusion length in the base and absorption length, respectively. $k$ is a scaling factor. In this equation, $W_{de}, L_{eff}$ and $k$ are the unknowns, which can be determined by a simple iterative process. It begins with a reasonable guess of the diffusion length. With this the slope and intercept of the $\ln(IQE(1+L_a/L_{eff}))$ vs. $1/L_a$ plot provides the value of $W_{de}$ and $k$ respectively. Based on the error present, the value of $L_{eff}$ is adjusted and another set of $W_{de}$ and $k$ are calculated. This iterative processes continues until it converges on the solution of the equation. Once all the parameters are obtained, they can be used for calculation of the individual losses. The dead layer thickness is valid for the shorter wavelengths. Thus, the total loss in the emitter can be divided into two parts,

$$A_{e,I}(\lambda) = 1 - IQE(\lambda)\cdot\left(1-\frac{L_a(\lambda)}{L_{eff}}\right) \quad (6a)$$

$$A_{e,II}(\lambda) = 1 - \exp\left(-\frac{W_{de}}{L_a(\lambda)}\right) \quad (6b)$$

The remaining parasitic losses can be attributed to the base and rear surface loss. $J_{SC}$ is an important parameter in determining cell performance. It can be determined from the following equation.

$$J_{sc} = e\int_{365\,nm}^{1280\,nm} E\,QE(\lambda)\phi_{in}(\lambda)\,d\lambda \quad (7)$$

Here, $e$ is the charge of an electron. The limits of the integrals are taken from 365 nm to 1280 nm, because this is the range of light the LEDs in the FlashQE emit.

The non-destructive cTLM measurements were performed using a BrightSpot Automation *ContactSpot-PRO* system. The contact resistivity ($\rho_C$) and emitter sheet resistance ($R_{sheet}$) were extracted from the cTLM measurements using the technique in [6]. The $\rho_C$ and $R_{sheet}$ of c-Si solar cells is traditionally measured using TLM measurements, wherein the test structures are created by isolating strips of a cell (e.g., laser scribe then cleave) or by fabricating special test structures within a wafer. The former can be applied to industrial cells, but is destructive, and the latter is non-destructive, but cannot be used on industrial cells. The non-destructive technique used in this work relies on hiding circular TLM (cTLM) structures within the busbars of the cells, and therefore does not result in any additional shading of the cell, so it can be used on finished solar cells.

## III. RESULTS AND DISCUSSION

Illuminated *I-V*, Suns-$V_{OC}$, PL, high speed EQE + *R*, and cTLM measurements were all carried out for the ≈400 finished solar cells coming from the same production line. Histograms of some of the various parameters determined from this combination of measurements are shown in Figure 3. The distributions of the parameters can help quantify variance due to materials and processes. It can also provide insight into what loss mechanisms are limiting cell performance and help prioritize efforts to optimize performance of the production line. For example, histograms of the loss parameters determined from the EQE + *R* data show that loss in the bulk/rear of the cell ($J_{loss-b}$) is the primary limiting factor for $J_{SC}$, followed by front surface reflectance ($J_{R-f}$).

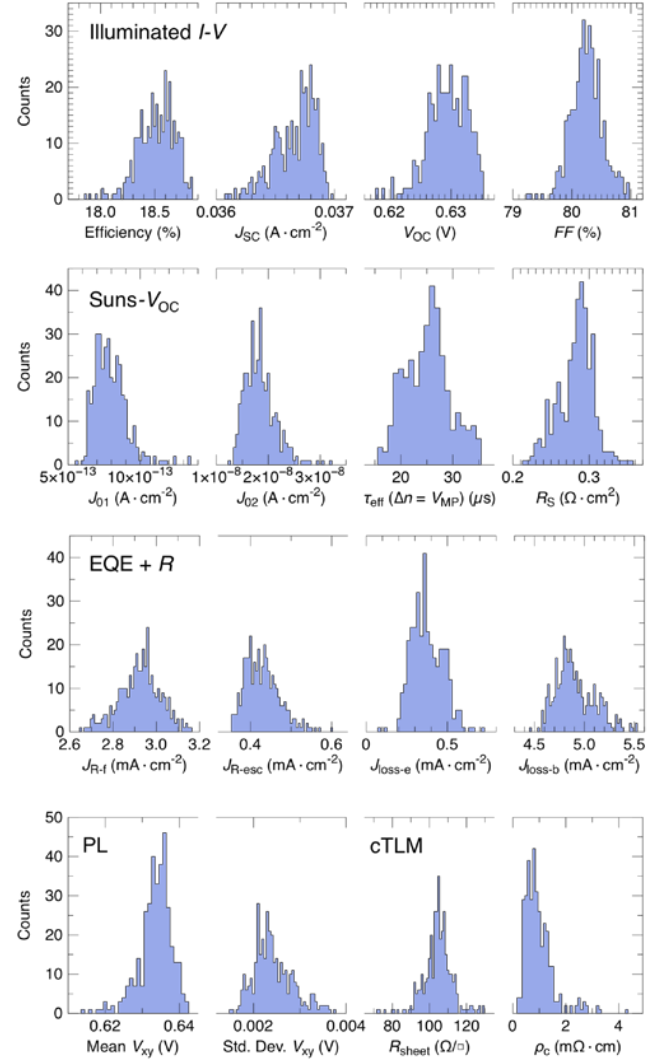

Fig. 3. Histograms of the various parameters calculated from the illuminated *I-V*, Suns-$V_{OC}$, PL, high speed EQE + *R*, and cTLM measurements. Note, $R_S$ is determined from the difference between the *I-V* and Suns-$V_{OC}$ curves.

Doing all these measurements for a high volume cell group has provided an unique opportunity to investigate the proper causes behind losses in solar cells and the relationships among different performance parameters. The 1 sun open-circuit PL and $J_{SC}$ image of three different cells are illustrated in Figure 4, a good cell, an average cell, and a poorly performing cell, in terms of efficiency. For each cell, the positions and patterns of these localized areas of poor performance are virtually identical in both images due to the known reciprocity EQE and PL.

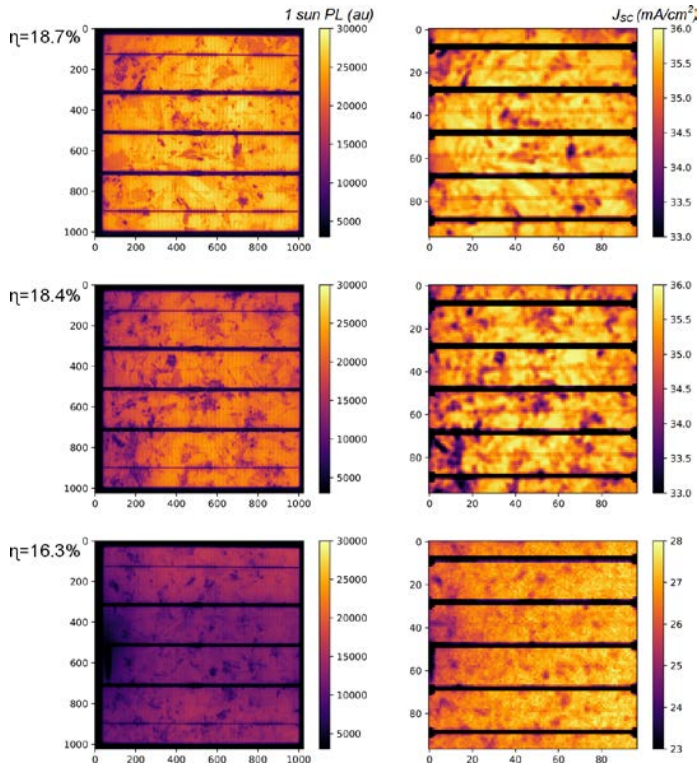

Fig. 4. PL images and spatially-resolved $J_{SC}$ of three different cells, a good, average, and a poorly performing cell - the efficiency values for each are given on the left hand side of the figures. Note, the poorly performing cell has a different $J_{SC}$ scale.

Efficiency is the most important parameter in solar cell production. The big data set coming from the $I$-$V$, Suns-$V_{OC}$, cTLM, PL and FlashQE measurements of the ≈400 cells was investigated to find out a comprehensive picture of how different parameters affect the cell efficiency. The efficiency from the $I$-$V$ measurements was compared with all other measured data. As depicted in the figure 5, a strong linear relationship is observed between the efficiency and $J_{01}$ loss. A similar, but relatively weaker relationship is noticed between efficiency and $J_{02}$.

As expected, efficiency showed a linear dependence on the spatially-averaged $J_{SC}$. The strong linear dependence on the spatial standard deviation of $J_{SC}$ shows a quantitative relationship between efficiency lost and within-wafer EQE variation, which in this case is driven by bulk and rear recombination (see the correlation between efficiency and $L_{eff}$). The mean PL intensities at 0.1 sun and 1 sun and the mean $V_{xy}$ derived from the PL images all show a strong correlation to efficiency. Interestingly, no correlation between efficiency and standard deviation of $V_{xy}$ is observed.

Another method of analyzing the data is to evaluate correlations between the numerous parameters measured with these different techniques, with a focus on relationships between parameters that can actually provide useful insight into manufacturing. For example, Figure 6(a) shows which region of the device is limiting the overall $\tau_{eff}$ of the cells. Here, the current loss in the bulk and rear of the cell ($J_{loss-b}$) shows a correlation to $\tau_{eff}$, whereas the current loss due to parasitic absorption in the SiN$_x$ ARC and recombination in the emitter ($J_{loss-e}$) does not. Additionally, and the effective base diffusion length ($L_{eff}$) also has a strong correlation to $\tau_{eff}$, further indicating that that recombination in the bulk and rear of the cell is limiting $\tau_{eff}$ and not recombination in the emitter. This makes sense considering these are all multi Al-BSF cells. Another example uses the Suns-$V_{OC}$ and cTLM measurements to separate three components of $R_S$. In Figure 6(b), the lumped cell $R_S$ is correlated to $R_{sheet}$, (cTLM), $\rho_b$ (Suns-$V_{OC}$), and $\rho_c$ (cTLM). The strength of the correlations indicates that the $R_{sheet}$ of the emitter has the strongest influence, followed by the $\rho_b$ of the wafer next. The $\rho_c$ of these specific cells is too low to significantly affect the overall $R_S$.

A correlation between $J_{loss-e}$ and the emitter $R_{sheet}$ is noticed in Figure 6(c). Another similar correlation is noticed between $R_{sheet}$ and dead layer thickness. Both $J_{loss-e}$ and dead layer thickness was derived from the FlashQE data and account for losses in the emitter, while $R_{sheet}$ was measured using cTLM method.

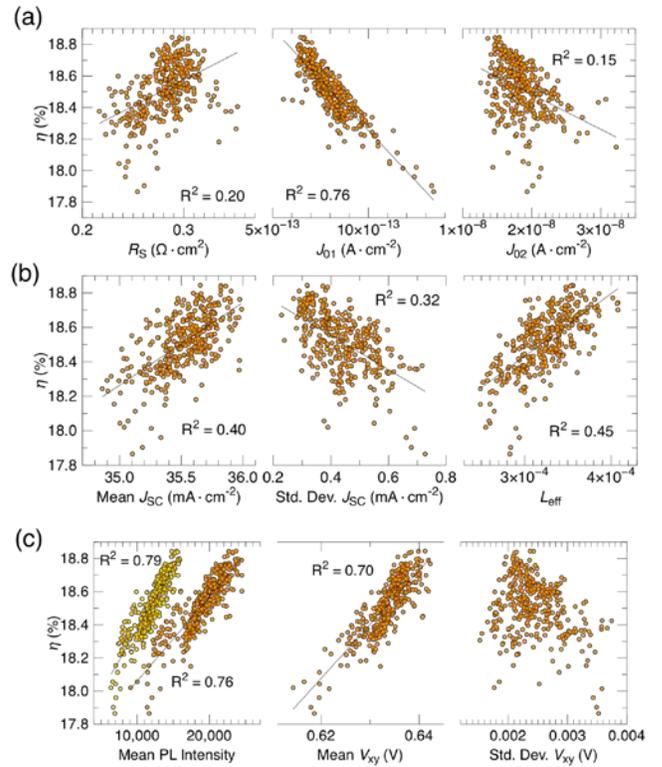

Fig. 5. Correlations between the cell efficiency and various parameters, including: (a) $R_S$, $J_{01}$, and $J_{02}$ measured using $I$-$V$ and Suns-$V_{OC}$; (b) the mean $J_{SC}$, within-cell standard deviation of $J_{SC}$, and $L_{eff}$ measured with EQE + R; and (c) the mean PL intensity at 0.1 sun (yellow) and 1 sun (orange), mean $V_{xy}$, and within-wafer standard deviation of $V_{xy}$ measured using PL imaging.

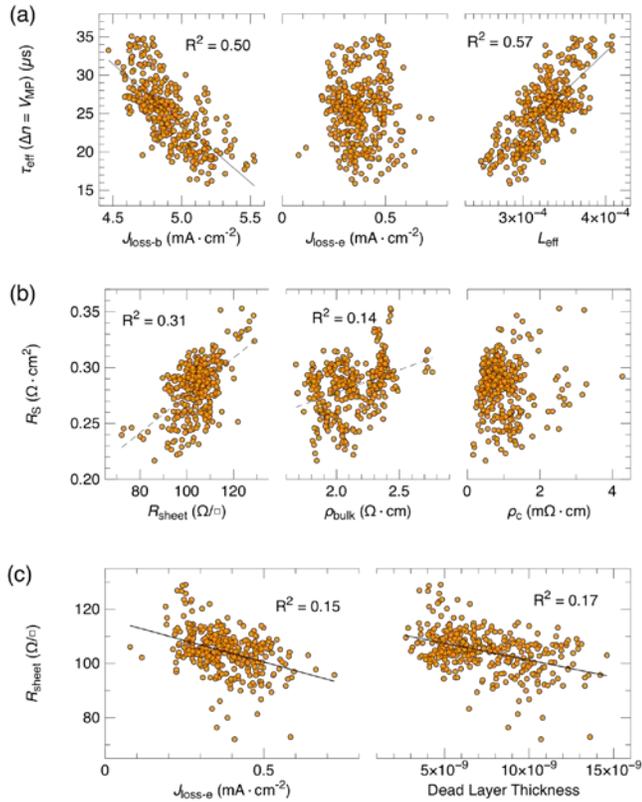

Fig. 6. (a) Correlations between $\tau_{eff}$ at $V_{MP}$, determined from Suns-$V_{OC}$, and $J_{loss-b}$, $J_{loss-e}$, and $L_{eff}$, all determined from the EQE data. (b) Correlations between $R_S$, determined from the difference in the $I$-$V$ and Suns-$V_{OC}$ curves, and $R_{sheet}$ (cTLM), $\rho_b$ (Suns-$V_{OC}$), and $\rho_c$ (cTLM). (c) Correlations between $R_{sheet}$, determined from cTLM, and both $J_{loss-e}$ and the dead layer thickness, both determined from the EQE data.

## IV. CONCLUSIONS

Access to larger sets of device parameters can provide useful insight into specific loss mechanisms and be used for process control and specific defects can be identified, classified, and their impact quantified. In this work, we observed a variety of notable trends. Spatial variation of $J_{SC}$ within a cell results in lower efficiency, whereas spatial variation in $V_{OC}$ did not. $J_{01}$ recombination had a stronger impact on cell efficiency than $R_S$. Carrier lifetime has strong correlation with base-loss and diffusion length, but no correlation with emitter-loss. This shows that this group is limited by recombination in the bulk and rear. $R_S$ is primarily influenced by emitter $R_{sheet}$, followed by the wafer $\rho_{bulk}$. No correlation was found between $R_S$ and $\rho_c$. $R_{sheet}$ has strong correlation with current loss in the emitter and the dead layer thickness, calculated from the QE at short wavelengths.

ACKNOWLEDGEMENTS

This work is supported by the U.S. Department of Energy SunShot Initiative under grant number DE-EE-0008155.